\documentclass[twocolumn,prb,aps,floats,floatfix,showpacs,amsmath,amssymb,citeautoscript]{revtex4}
\usepackage{graphicx}
\usepackage{bm}
\usepackage{times}

\def\Xint#1{\mathchoice
   {\XXint\displaystyle\textstyle{#1}}%
   {\XXint\textstyle\scriptstyle{#1}}%
   {\XXint\scriptstyle\scriptscriptstyle{#1}}%
   {\XXint\scriptscriptstyle\scriptscriptstyle{#1}}%
   \!\int}
\def\XXint#1#2#3{{\setbox0=\hbox{$#1{#2#3}{\int}$}
     \vcenter{\hbox{$#2#3$}}\kern-.5\wd0}}

\def\dashint{\Xint-}

\begin{document}

\newcommand{\be}{\begin{equation}}
\newcommand{\ee}{\end{equation}}
\newcommand{\bea}{\begin{eqnarray}}
\newcommand{\eea}{\end{eqnarray}}
\newcommand{\beann}{\begin{eqnarray*}}
\newcommand{\eeann}{\end{eqnarray*}}
\newcommand{\bma}{\begin{array}{cc}}
\newcommand{\ema}{\end{array}}
\newcommand{\fr}{\frac}
\newcommand{\ra}{\rangle}
\newcommand{\la}{\langle}
\newcommand{\li}{\left}
\newcommand{\re}{\right}
\newcommand{\ri}{\right}
\newcommand{\uarr}{\uparrow}
\newcommand{\darr}{\downarrow}
\newcommand{\df}{\stackrel{\rm def}{=}}
\newcommand{\nn}{\nonumber}
\newcommand{\dpl}{\displaystyle}

\newcommand{\alp}{\alpha}
\newcommand{\sig}{\sigma}
\newcommand{\eps}{\epsilon}
\newcommand{\xsi}{\xi}
\newcommand{\lam}{\lambda}
\newcommand{\ny}{\nu}

\newcommand{\logicup}{+}
\newcommand{\logicdown}{-}
\newcommand{\logicexc}{e}
\newcommand{\exppl}{{e^{i \beta}}}
\newcommand{\expmi}{{e^{-i \beta}}}
\newcommand{\tpm}{t^\prime e^{i\beta}/3}
\newcommand{\tpn}{t^\prime e^{-i\beta}/3}
\newcommand{\tppm}{t^{\prime\prime} e^{i\beta}/3}
\newcommand{\tppn}{t^{\prime\prime} e^{-i\beta}/3}

\title{Improving Intrinsic Decoherence in Multiple-Quantum-Dot Charge Qubits}

\author{Martina Hentschel,$^1$ Diego C. B. Valente,$^2$ Eduardo
R. Mucciolo,$^2$ and Harold U. Baranger$^3$}

\affiliation{$^1$Max-Planck-Institut f\"ur Physik Komplexer Systeme,
N\"othnitzer Stra\ss e 38, Dresden, Germany}

\affiliation{$^2$Department of Physics, University of Central Florida,
P.O. Box 162385, Orlando, Florida 32816-2385, USA,}

\affiliation{$^3$Department of Physics, Duke University, P.O. Box
90305, Durham, North Carolina 27708-0305, USA}

\date{\today; arXiv:0705.3923}

\begin{abstract}
We discuss decoherence in charge qubits formed by multiple lateral
quantum dots in the framework of the spin-boson model and the
Born-Markov approximation. We consider the intrinsic decoherence
caused by the coupling to bulk phonon modes. Two distinct quantum dot
configurations are studied: (i) Three quantum dots in a ring geometry
with one excess electron in total and (ii) arrays of quantum dots
where the computational basis states form multipole charge
configurations. For the three-dot qubit, we demonstrate the
possibility of performing one- and two-qubit operations by solely
tuning gate voltages. Compared to a previous proposal involving a
linear three-dot spin qubit, the three-dot charge qubit allows for
less overhead on two-qubit operations. For small interdot tunnel
amplitudes, the three-dot qubits have $Q$ factors much higher than
those obtained for double-dot systems. The high-multipole dot
configurations also show a substantial decrease in decoherence at low
operation frequencies when compared to the double-dot qubit.
\end{abstract}


\maketitle
\section{Introduction}

The realization of a solid-state qubit based on familiar and highly
developed semiconductor technology would facilitate scaling to a
many-qubit computer and make quantum computation more
accessible.\cite{QCreviews} The earliest proposal of a quantum dot
qubit relied on the manipulation of the spin degree of freedom of a
single confined electron.\cite{loss98} An attractive point of that
proposal is the large spin decoherence time characteristic of
semiconductors; a drawback is that it requires local control of
intense magnetic fields. As an alternative, a spin-based logical qubit
involving a multiple quantum dot setup and voltage-controlled exchange
interactions was devised,\cite{divincenzo00} but at the price of
considerable overhead in additional operations.

While spin qubits remain promising in the long term -- note in
particular several recent experimental
advances\cite{petta05,koppens06} as well as further theoretical
development of multiple-quantum dot spin qubits
\cite{LevyWeinsteinHellberg} -- charge-based qubits in quantum dots,
in analogy to superconducting Cooper-pair box
devices,\cite{nakamura,vion,blais,koch} are also worthy of
investigation. Employing the charge degree of freedom of electrons
rather than their spin brings a few important practical advantages: No
local control of magnetic fields is required and all operations can be
carried out by manipulating gate voltages. The simplest realization of
a charge qubit is a double quantum dot system with an odd number of
electrons.\cite{blick00, tanamoto00, fedichkin04, brandes02, wu04,
sergueietaldoubledot05, Hu05} One can view this system as a double
well potential: The unpaired electron moves between the two wells
(i.e., quantum dots) by tunneling through the potential barrier. The
logical states $|0\rangle, |1\rangle$ correspond to the electron being
on the left or right. The barrier height determines the tunneling rate
between the dots and can be adjusted by a gate voltage. The resulting
bonding and antibonding states can also be used as the computational
basis. Recently, three groups have implemented the double-dot charge
qubit experimentally.\cite{hayashi03,petta04,gorman05}

Charge qubits are susceptible to various decoherence mechanisms
related to charge motion. Strong damping of coherent oscillations was
observed in all quantum dot experimental
setups,\cite{hayashi03,petta04,gorman05} with quality factors in the
range $3$-$10$. Note that a change in the state of the qubit involves
electron motion between quantum dots, which can in general couple very
effectively to external degrees of freedom such as phonons, charge
traps in the substrate, and electromagnetic environmental
fluctuations. These noise sources lead to decoherence times much
shorter than those observed in spin qubit systems. Thus, one is
tempted to try to find new setups where oscillations between qubit
states involve a minimum amount of charge motion. For instance, in
qubits based on multiple quantum dots one can pick logical states
where charge is homogeneously distributed in space. Another approach
is to create a multiple dot structure with symmetries that forbid
coupling to certain environmental modes within the logical
subspace.\cite{kempe} A recent attempt along this direction is found
in Ref.\,\onlinecite{oi05}.

In this paper we argue that it is not generally possible to avoid
decoherence in multiple-quantum-dot charge qubits by simple
geometrical constructions. The spreading of charge uniformly over a
multiple-quantum-dot logical qubit does not avoid
decoherence. However, the coupling to bosonic environmental modes,
such as phonons and photons, can be very substantially attenuated in
some circumstances.

\begin{figure}[t]
\includegraphics[width=5cm]{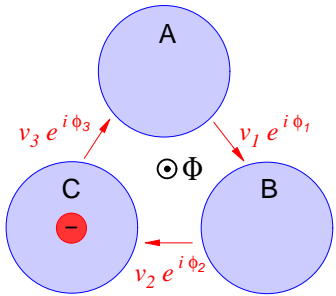}
\caption{(Color online) Schematic illustration of a three-quantum-dot
qubit with only one extra, unpaired electron. The external tuning
parameters are the strength of the tunneling couplings ($v_1$, $v_2$,
and $v_3$) and the magnetic flux $\Phi = \phi_1+\phi_2+\phi_3$ through
the qubit. The latter is used solely to define the working point of
the qubit.}
\label{fig_3dotqubit}
\end{figure}

In order to demonstrate these assertions, we analyze in detail two
prototypical extensions of the double-quantum dot charge qubit. We
first consider a qubit consisting of three quantum dots forming a
ringlike structure and only one extra electron, as shown in
Fig.~\ref{fig_3dotqubit}. Multiple-quantum-dot qubits with a ringlike
structure resemble a proposal by Kulik {\it et al.} \cite{kulik} to
use persistent current states in metallic rings for quantum
computation. Unlike the double-dot qubit case, the ground state in a
three-dot qubit can be truly degenerate with corresponding wave
functions having a uniform charge distribution. At first, this raises
the hope that decoherence mechanisms involving charge inhomogeneities
(such as phonons or charge traps) would be inhibited due to mutual
cancellations. However, we shall see below that the computational
basis states can be distinguished by phonon and electromagnetic baths
through the electron phase variations along the ring. That, in turn,
leads to dephasing and decoherence. This problem is intrinsic to all
quantum-dot-based charge qubits. Nevertheless, the $Q$ factor in these
three dot qubits can be 1-2 orders of magnitude larger than in the
corresponding double-dot qubits, a substantial improvement in
coherence.

Second, we show that planar quantum dot arrays in the form of
high-order multipoles can be more efficient in reducing the coupling
to acoustic phonons in multiple-quantum dot qubits. Our work extends
and analyzes in detail a recent proposal to create a decoherence-free
subspace with charge qubits.\cite{oi05}

While it is well known that condensed-matter environments tend to
produce time and spatial correlations in their interaction with
qubits,\cite{nonmarkovian} here we assume
that the Markov approximation provides reasonable estimates
of decoherence rates. In particular, we employ the Redfield
equations in the weak-coupling, Born-Markov approximation to describe
the time evolution of the reduced density matrix of the qubit
system.\cite{argyres64}

The paper is organized as follows. In Sec.~\ref{sec:threedot} we study
in detail a three-dot charge qubit. Single- and two-qubit operations
are presented, as well as the coupling to a bosonic bath. We consider
in detail the particular case of acoustic piezoelectric phonons, which
is relevant to III-V semiconductor materials at low temperatures. Also
in Sec.~\ref{sec:threedot} we evaluate decoherence and energy
relaxation rates using the Redfield equation formalism. In
Sec.~\ref{sec:multipledots} we present a multiple-quantum dot logical
qubit structure that minimizes the coupling to environmental modes
which couple to charge. We also analyze phonon decoherence in these
systems and compare $Q$ factors with those obtained with double-dot
charge qubits. Finally, our conclusions are presented in
Sec.~\ref{sec:conclusions}. The Appendix contains mathematical details
of the two-qubit operation of the three-dot charge qubit of
Sec.~\ref{sec:threedot}. Throughout this paper we assume $\hbar=1$ and
$k_B=1$.

\section{The three-dot charge qubit}
\label{sec:threedot}

\begin{figure}[t]
\includegraphics[width=8.5cm]{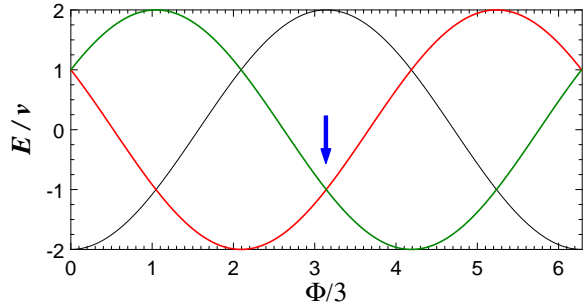}
\caption{(Color online) Eigenenergies of the three-dot qubit as
function of the magnetic flux. The working point at $\Phi/3 =\phi =
\pi$ per bond is indicated by the arrow. At this point, clockwise and
counterclockwise persistent current states are degenerate, and the
charge distribution is homogeneous throughout the space spanned by the
computational basis.
\label{fig_eigenenergies}}
\end{figure}

A simple example of a multiple-dot qubit with charge delocalization
consists of three quantum dots in a ringlike geometry, as shown in
Fig.~\ref{fig_3dotqubit}. In practice, this system is created by
laterally confining electrons in a two-dimensional plane; the
confinement is electrostatic, controlled through electrodes sitting
above the plane. Consider gate voltages on the electrodes such that
the three dots share one excess, unpaired electron, while all
configurations with a different number of excess electrons become
energetically inaccessible due to the large charging energy of the
dots.\cite{Hawrylak07} The spin degree of freedom is not relevant for
our discussion and electrons will be assumed spinless unless otherwise
specified. Thus, the system lives in a three-dimensional Hilbert
space. The electron can hop between dots through tunneling. The
tunneling matrix elements and the on-site energies are controlled by
the gate voltages. As will be clear shortly, it is convenient to apply
a weak magnetic field perpendicular to the plane containing the dots.

The three natural basis states place the electron on dot A, B, or C:
\begin{equation}
|A \rangle = c_A^\dagger\, | {\rm vac.} \rangle, \ \ \
|B \rangle = c_B^\dagger\, | {\rm vac.} \rangle, \ \ \
|C \rangle = c_C^\dagger\, | {\rm vac.} \rangle,
\end{equation}
where $c_\alpha^\dagger$ are creation operators and $|{\rm
vac.}\rangle$ is a reference state where all dots have an even number
of electrons. In this basis, the Hamiltonian takes the matrix form
\begin{equation}
\label{eq:Hamilton}
H = \left( \begin{array}{ccc} E_A & -v_1\, e^{i\phi_1} & -v_3\,
e^{-i\phi_3} \\ -v_1\, e^{-i\phi_1} & E_B & -v_2\, e^{i\phi_2} \\
-v_3\, e^{i\phi_3} & -v_2\, e^{-i\phi_2} & E_C
\end{array} \right),
\end{equation}
where $E_A,E_B$, and $E_C$ are the on-site energies, $v_i$ are the
tunneling strengths between pairs of quantum dots, and $\phi_1 \!+\!
\phi_2 \!+\! \phi_3 = \Phi$ is the total magnetic flux through the
ring. Let us specify the qubit by setting $v_1 \!=\! v_2 \!=\!
v_3\!\equiv\! v\!>\!0$, $E_A \!=\! E_B \!=\! E_C \!\equiv\! 0$, and
$\phi_1 \!=\! \phi_2 \!=\! \phi_3 \!=\! \Phi/3 \!\equiv\! \pi$. In
this configuration, two degenerate eigenstates $|+\rangle$ and
$|-\rangle$ have the lowest energy, $E_{\pm} = -v$
(Fig.~\ref{fig_eigenenergies}). They carry clockwise and
counterclockwise persistent currents and form the computational
basis.\cite{note_holestates} The third, excited, eigenstate $|T\rangle$ has energy
$E_{\logicexc} = 2v$ and is current-free. The eigenvectors are
\bea |T\rangle & = & \frac{1}{\sqrt{3}} \left( |A\rangle + |B\rangle +
|C\rangle \right), \\
\label{logicpl}
|+\rangle & = & \frac{1}{\sqrt{3}} \left( |A\rangle + e^{i \beta}
|B\rangle + e^{-i \beta} |C\rangle \right), \\
\label{logicmi}
|-\rangle & = & \frac{1}{\sqrt{3}} \left( |A\rangle + e^{-i \beta}
|B\rangle + e^{i \beta} |C\rangle \right), 
\eea
with $\beta= 2 \pi/3$. Clearly, the charge distribution is spatially
uniform for all three states.

It is worth noting that the topology of the three-dot qubit and its
use of persistent currents of opposite direction as logical states
closely resemble the Josephson persistent current qubit studied in
Ref. \,\onlinecite{mooij1999} or the proposed atomic Josephson
junction arrays. \cite{zoller2003} However, the similarities stop here
as the underlying physics is very different. We will focus our
discussion on the quantum dot charge qubit case only.

\subsection{Single-qubit operations}
\label{sec:threedotops_single}

In order to be able to perform quantum gate operations, we have to
allow for deviations from the degeneracy point. This is done by
varying the tunneling coupling and/or the magnetic flux. It is
convenient to introduce the (small) parameters $\delta_1$, $\delta_2$,
$\delta_3$, and $\varphi$ such that $v_1 \!=\!  v + \delta_1$, $v_2
\!=\!  v + \delta_2$, $v_3 \!=\!  v + \delta_3$, and $\varphi \ll 1$
with $\varphi \!=\!  \Phi - 3\pi$. To linear order and using a
$\{|T\rangle,|+\rangle,|-\rangle\}$ basis, we find that the
Hamiltonian expanded around the degeneracy point can be written as
\begin{widetext}
\begin{equation}
\label{eq:H3dot}
H = \left( \begin{array}{ccc} 2v + \frac{2}{3}\left( \delta_1 +
     \delta_2 + \delta_3 \right) & - \frac{1}{3}\left( \delta_1 e^{-i
     \beta} + \delta_2 + \delta_3 e^{i \beta} \right) & -
     \frac{1}{3}\left( \delta_1 e^{i \beta} + \delta_2 + \delta_3
     e^{-i \beta} \right) \\ - \frac{1}{3}\left( \delta_1 e^{i \beta}
     + \delta_2 + \delta_3 e^{-i \beta} \right) & -v -v
     \varphi/\sqrt{3} -\frac{1}{3}\left( \delta_1 + \delta_2 +
     \delta_3 \right) & \frac{2}{3}\left( \delta_1 e^{-i \beta} +
     \delta_2 + \delta_3 e^{i \beta} \right) \\ - \frac{1}{3}\left(
     \delta_1 e^{-i \beta} + \delta_2 + \delta_3 e^{i \beta} \right) &
     \frac{2}{3}\left( \delta_1 e^{i \beta} + \delta_2 + \delta_3
     e^{-i \beta} \right) & -v + v \varphi/\sqrt{3} -\frac{1}{3}\left(
     \delta_1 + \delta_2 + \delta_3 \right)
\end{array} \right).
\end{equation}
\end{widetext}
The computational subspace corresponds to the lower-right
$2\!\times\!2$ block. Evidently, we stay within the computational
subspace as long as $\delta_1=\delta_2=\delta_3$. However, this also
implies that there is no coupling between the computational basis
states $|+\rangle$ and $|-\rangle$. For $\delta_1 e^{i \beta} +
\delta_2 + \delta_3 e^{-i \beta} \neq 0$, coupling within the
computational subspace is possible, but there is a finite probability
of leaking out into the state $|T\rangle$. The leakage can be kept
small as long as $v\gg |\delta_{1,2,3}|$. Alternatively, one can
incorporate the third level into the single-qubit operations, as in
Ref.\,\onlinecite{kulik}. For the following case study, we assume that
the leakage from the computational subspace is negligible.

Using the Pauli matrices $\sigma_1$, $\sigma_2$, and $\sigma_3$, as
well as the identity matrix $\sigma_0$, we can express the Hamiltonian
in the computational basis in terms of a pseudospin in a
pseudomagnetic field $\vec{h}$ plus a constant,
\begin{equation}
H_S = E_0\, \sigma_0 + h_x\, \sigma_1 + h_y\, \sigma_2 + h_z\,
\sigma_3,
\end{equation}
where $E_0 = -v - (\delta_1+\delta_2+\delta_3)/3$ and
\begin{eqnarray}
h_x & = & \frac{2}{3} \left( \delta_2 - \frac{\delta_1 +
\delta_3}{2} \right), \\ h_y & = & \frac{\delta_1 -
\delta_3}{\sqrt{3}}, \\ h_z & = & - v \varphi/\sqrt{3}.
\label{pseudoBz}
\end{eqnarray}
We only need to vary two out of the three pseudomagnetic field
components in order to perform single-qubit operations. Thus, we can
operate the qubit at constant magnetic flux (and set $\varphi \!=\!
0$, $h_z \!=\! 0$) and vary only the $\delta_i$ via gate voltages. If
we furthermore fix the coupling $v_2 \!\equiv\! v$, $\delta_2 \!=\! 0
$, we find that the qubit is controlled by the sum and difference of
the variation of two intra-qubit couplings, $h_x \propto (\delta_1 +
\delta_3)$ and $h_y \propto (\delta_1 - \delta_3)$, that can be
adjusted by tuning the respective gate voltages around the symmetry
point.

\subsection{Two-qubit operations}
\label{sec:threedotops_two}

\begin{figure}[b]
\includegraphics[width=8.0cm]{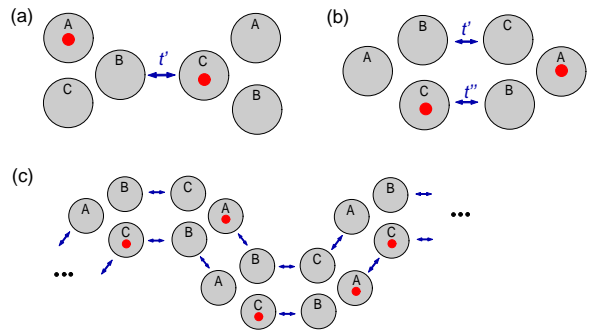}
\caption{(Color online) Possible implementations of two-qubit gates
using three-dot qubits. (a) Coupling via a single dot (tip-tip
geometry); (b) coupling via two dots (base-base geometry). (c) A
possible implementation of a qubit chain in the base-base
configuration.}
\label{fig_2qubitoper}
\end{figure}

In order to perform two-qubit operations, such as the SWAP or CNOT
gate, we have to couple two three-dot qubits (called I and II
hereafter). In principle, this can be done in either a tip-to-tip or
base-to-base coupling scheme, as shown in
Fig.~\ref{fig_2qubitoper}. Since the number of excess electrons in the
composite system is equal to two, states where two electrons occupy
the same qubit have to be included in the basis of the two-qubit
Hilbert space. 
The basis of the two-qubit Hilbert space reads thus
\begin{eqnarray}
\label{eq:1}
| 1\rangle & = & |\logicup\rangle_{\rm I}\, |\logicup\rangle_{\rm II},
\label{2qubitbasis1}\\
\label{eq:2}
|2\rangle & = & |\logicup\rangle_{\rm I}\,|\logicdown\rangle_{\rm II},
 \\
\label{eq:3}
|3\rangle & = & |\logicdown\rangle_{\rm I}\, |\logicup\rangle_{\rm
 II}, \\
\label{eq:4}
|4\rangle & = & |\logicdown\rangle_{\rm I}\, |\logicdown\rangle_{\rm
II}, \\ |5\rangle & = & c^\dagger_{A\rm I}\, c^\dagger_{B\rm I}\,
|{\rm vac.} \rangle,\\ |6\rangle & = & c^\dagger_{A\rm I}\,
c^\dagger_{C\rm I}\, |{\rm vac.}  \rangle, \\ |7\rangle & = &
c^\dagger_{B\rm I}\, c^\dagger_{C\rm I}\, |{\rm vac.} \rangle, \\
|8\rangle & = & c^\dagger_{A\rm II}\,c^\dagger_{B\rm II}\, |{\rm vac.}
\rangle, \\ |9\rangle & = & c^\dagger_{A\rm II}\,c^\dagger_{C\rm II}\,
|{\rm vac.}  \rangle, \\ |10\rangle & = & c^\dagger_{B\rm
II}\,c^\dagger_{C\rm II}\, |{\rm vac.}
\label{2qubitbasis10}
\rangle.
\end{eqnarray}
Here, two types of states have been neglected: First, states with
double occupancy of a single dot since the charging energy is assumed
to be very large. Second, although the $|T\rangle_{\rm I}$ and
$|T\rangle_{\rm II}$ states couple to the double-occupied states
$|5\rangle$ to $|10\rangle$ through the inter-qubit hopping terms,
they are gapped by an energy of order $v$, which is assumed much
larger than the effective two-qubit interaction
amplitude $t^{\prime\, 2}/U_i$ (see below). Therefore, they were not
included in the two-qubit Hilbert subspace.\cite{technicalnote}

The Hamiltonian for the inter-qubit interaction in the tip-tip setting
shown in Fig.~\ref{fig_2qubitoper}(a) reads
%
\begin{equation}
H_{\rm I-II}^{\rm tip} = -t^\prime ( c^\dagger_{B\rm I}\,
c^{}_{C\rm II} + c^\dagger_{C\rm II}\, c^{}_{B\rm I}).
\label{hswaptiptip}
\end{equation}
Similarly, the base-base coupling presented in
Fig.~\ref{fig_2qubitoper}(b) is governed by the Hamiltonian (see also
the Appendix)
\begin{equation}
H_{\rm I-II}^{\rm base} = -t^\prime ( c^\dagger_{B\rm I}\,
c^{}_{C\rm II} + c^\dagger_{C\rm II}\, c^{}_{B\rm I} )
-t^{\prime \prime} ( c^\dagger_{C\rm I}\, c^{}_{B\rm II} +
c^\dagger_{B\rm II}\, c^{}_{C\rm I} ),
\label{hswapbasebase}
\end{equation}
where we have chosen the gauge for the vector potential associate to
the perperdincular magnetic field to be parallel to the inter-qubit
tunneling paths. We assume that the inter-qubit tunneling amplitudes
$t^\prime$ and $t^{\prime \prime}$ satisfy $0 < t^\prime, t^{\prime
\prime} \ll v \ll U_i$, where $U_i$ is the inter-dot charging or
capacitive coupling energy (i.e., the change in the energy of one dot
when an electron is added to one of the neighboring dots). In other
words, the capacitive coupling between dots must be sufficiently
strong so that states with two or zero excess electrons in a qubit are
forbidden. Due to the proximity between dots of neighboring qubits,
some small inter-qubit capacitive coupling will also exist. Although
we will neglect such coupling in the discussion below, these
additional charging energies can be included without substantially
modifying our results. In particular, we note that the inter-qubit
capacitive coupling does not interfere with single-qubit
operations. Note also that the presence of a magnetic flux requires
the dots A, B, and C to be always arranged in a clockwise order.

Next, the large charging energy separation between the
single-occupancy states $|1\rangle$ to $|4\rangle$ and the
double-occupancy states $|5\rangle$ to $|10\rangle$ allows us to
separate the two-qubit computational subspace from the rest of the
Hilbert space. In order to do so, we use a Schrieffer-Wolff
transformation,\cite{schriefferwolfftrafo} which amounts to a
second-order perturbative expansion of the effective Hamiltonian in
the ratio of the inter-qubit tunneling magnitude to the charging
energy.
To this end we insert the expressions for $|\logicup\rangle$ and
$|\logicdown\rangle$ from Eqs.~(\ref{logicpl})-(\ref{logicmi}) into
Eqs.~(\ref{eq:1})-(\ref{eq:4}) and express the computational basis
states $|1\ra$ to $|4\ra$ in terms of creation operators acting on the
vacuum state.
Further, using the basis vectors in Eqs.~(\ref{eq:1}) to
(\ref{2qubitbasis10}), one can easily compute the full six-dot
Hamiltonian in the basis of states $|1\ra$ to $|10\ra$.
Noting that one can obtain the tip-tip Hamiltonian from the expression
for the base-base case by setting $t^{\prime\prime}=0$, we evaluate
the more general case of the base-base coupling, see
Fig.~\ref{fig_2qubitoper}b and Eq. (\ref{hswapbasebase}). The details
of the computation, i.e. the full matrix representation of this
Hamiltonian, as well as its reduction to the two-qubit computational
basis by performing the Schrieffer-Wolff transformation, are shown in
Appendix \ref{ReducedHamAppendix}. The result for the reduced
Hamiltonian takes a rather compact form which, for the tip-tip case,
reads
\begin{equation}
\label{eq:Hreduced1}
\tilde{H}^{\rm tip}_{\rm I-II} = -\frac{t^{\prime\, 2}}{9\, U_i}\,
\left(
\begin{array}{cccc} 4 & \expmi & \exppl & -2 \\ \exppl & 4 & -2\expmi
& \exppl \\ \expmi & -2 \exppl & 4 & \expmi \\ -2 & \expmi & \exppl &
4
\end{array} \right).
\end{equation}
Note that this reduced Hamiltonian acts on the subspace formed by the
states $\{|1\rangle, \ldots |4\rangle\}$ defined in Eqs.~(\ref{eq:1})
to (\ref{eq:4}). Up to the common prefactor $-t^{\prime\, 2}/(9\,
U_i)$, the eigenvalues of $\tilde{H}^{\rm tip}_{\rm I-II}$ are $E_1 =
0$, $E_2 = 4$, $E_3 = 6$, and $E_4 = 6$, with the respective
eigenvectors equal to
\begin{eqnarray}
|E_1 \rangle & = & \frac{1}{2} \Big( |1\rangle - e^{i\beta}
|2\rangle - e^{-i\beta} |3\rangle + |4\rangle \Big), \\ 
|E_2 \rangle & = & \frac{1}{2} \Big( |1\rangle + e^{i\beta} 
|2\rangle + e^{-i\beta} |3\rangle + |4\rangle \Big), \\ 
|E_3 \rangle & = & \frac{1}{\sqrt{2}} 
\big( |1 \rangle -  |4 \rangle \big), \\ |E_4 \rangle & = &
\frac{1}{\sqrt{2}} \Big( e^{i\beta} |2\rangle - e^{-i\beta}
|3\rangle \Big).
\end{eqnarray}

The critical question now is whether this setup permits a convenient
two-qubit operation, such as a full SWAP. It is straightforward to
show that the answer is positive, even in the simple tip-tip coupling
scheme. To see that, suppose we initialize the qubits in state
$|2\rangle$ and now search for the time $\tau$ after which the qubits
have evolved onto the (swapped) state $|3\rangle$ under the action of
$\tilde{H}^{\rm tip}_{\rm I-II}$. The square of the resulting
condition,
$ \left| \langle 3 | e^{-i \tilde{H}^{\rm tip}_{\rm I-II} \tau} | 2
\rangle \right| ^2 \equiv 1 $, is readily evaluated and yields $\tau_S
= \pi/2 \; [t^{\prime\, 2}/(9\, U_i)]^{-1}$ as the (shortest) time for
which the tip-tip coupling $t^{\prime}$ has to be turned on in order
to implement the SWAP gate.

For a comparison with the (linear) three-dot spin qubit scheme
proposed by DiVincenzo {\it et al.}, \cite{divincenzo00} let us
briefly discuss the implementation of the CNOT quantum gate. A CNOT
can be done straightforwardly using two $\sqrt{\rm SWAP}$ operations
(SWAP gates of duration $\tau_S/2$) and seven one-qubit gates
\cite{loss98,siewert}, e.g., by utilizing the scheme in
Ref.\,\onlinecite{siewert}. Consequently, we find that the realization
of one- and two-qubit operations for the present three-dot charge
qubit is considerably simpler than for the proposal by DiVincenzo et
al. where many more steps were necessary to implement a CNOT. One
reason is the complexity of the one-qubit rotations -- for the logical
spin-qubit, one-qubit operations alone require three spin exchange
interaction pulses. For the CNOT gate, this implies at least 19 pulses
with 11 different operation times. Compared to the 9 pulses needed for
the three-dot charge qubit, the practical advantages of the qubit and
computation scheme proposed here are evident.

\subsection{Coupling to a bosonic bath}
\label{sec:phonondecoherence}

The charge qubit couples to a variety of environmental degrees of
freedom. We study in particular the decoherence caused by gapless
bosonic modes that sense charge fluctuations in the dots, such as
phonons. We assume that all quantum dots couple to the same bath. The
Hamiltonian describing the non-interacting bosonic modes in this case
is
\begin{equation}
H_B \!=\! \sum_{\bf q} \omega_{\bf q}\, b_{\bf q}^\dagger b_{\bf q},
\end{equation}
with ${\bf q}$ denoting the boson linear momentum and $\omega_{\bf q}$
its dispersion relation. The coupling between the dots and the bosons
is assumed to be governed by the bilinear Hamiltonian \cite{spie}
\begin{equation}
H_{\rm dot-boson} = \sum_{\bf q} (\alpha_A N_A + \alpha_B N_B +
\alpha_C N_C)( b_{\bf q}^\dagger + b_{-\bf{q}} ),
\end{equation}
where $N_k$ is the number operator of the $k$th dot, and
\begin{equation}
\label{eq:alpha}
\alpha_k = \lambda_{\bf q}\, P_{\bf q}^{(k)}\, e^{i {\bf R}_k \cdot
{\bf q}} \;.
\end{equation}
Here, $\lam_{\bf q}$ represents the electron-boson coupling constant
and $P_{\bf q}^{(k)}$ and $\mathbf{R}_k$ are form factor and position
vector of the $k$th dot, respectively. Note that all geometrical
information is contained in the coefficients $\alpha_k$. Since we have
exactly one excess electron on the three-dot system, the constraint
$N_A + N_B + N_C = 1$ must be satisfied. Therefore, the system-bath
Hamiltonian in the basis $\{|A\rangle, |B\rangle, |C\rangle\}$ reads
\begin{equation}
H_{\rm SB} = \sum_{\bf q} \left( \begin{array}{ccc} \alpha_A & 0 & 0
\\ 0 & \alpha_B & 0 \\ 0 & 0 & \alpha_C \end{array} \right) (b_{\bf
q}^\dagger + b_{-{\bf q}}) \;.
\end{equation}
Projection of this Hamiltonian onto the subspace spanned by
$|\logicup\ra$ and $|\logicdown\rangle$ defined in
Eqs.~(\ref{logicpl}) and (\ref{logicmi}) constrains the coupling to
that subspace, yielding
\begin{widetext}
%
%
\begin{equation}
\label{eq:HredSB}
\tilde{H}_{\rm SB} = \frac{1}{3} \sum_{\bf q} \left[ \left( \alpha_A -
            \frac{\alpha_B + \alpha_C}{2} \right) { \sigma_1} -
            \frac{\sqrt{3}}{2} \left( \alpha_B - \alpha_C \right)
            {\sigma_2} \right] \left( b_{\bf q}^\dagger + b_{-{\bf q}}
            \right),
\end{equation}
\end{widetext}
where a term proportional to $\sigma_0$ has been dropped. The presence
of two terms with different functional dependence on $\mathbf{q}$
indicates the coupling to two bath modes, which will be denoted by the
indices $1$ and $2$ in the following. There would be a third bath
mode, proportional to $\sigma_3$, if the charge distribution were not
the same for the two logical states. \textit{The advantage of having a
homogeneous charge distribution for both states in the computational
basis, leading directly to the cancellation of this third mode of
decoherence, is evident here.} It is important to remark that charge
homogeneity can be achieved without the assumptions of homogeneous
tunneling or equal capacitances: as long as one can tune the gate
voltages in the quantum dots independently, one can arrange to have
one extra electron equally shared among the three dots.

It is convenient to rewrite the system-bath Hamiltonian in the
standard spin-boson form \cite{spinbosonmodel}
\begin{equation}
\tilde{H}_{\rm SB} \equiv K_1 \Phi_1 + K_2 \Phi_2 \:,
\end{equation}
where 
\begin{equation}
K_1 \equiv {\sigma_1}/6 \quad\mbox{and}\quad K_2 \equiv - {\sigma_2}/
2 \sqrt{3}
\end{equation}
describe the system part and the corresponding bath part is given by
$\Phi_{1,2} = \sum_q g^{(1,2)}_{\bf q} \left( b_{\bf q}^\dagger +
b_{-{\bf q}} \right)$, with
\begin{eqnarray}
\label{eq:g1}
g^{(1)}_{\bf q} & = & 2 \alpha_A - \alpha_B - \alpha_C, \\
\label{eq:g2}
g^{(2)}_{\bf q} & = & \alpha_B - \alpha_C \;.
\end{eqnarray}
Assuming all $P_{\bf q}^{(k)}$ to be the same, the following relations
among the $\alpha_k$ can be obtained:
\begin{eqnarray}
\alpha_A & = & \lambda_{\bf q} P_{\bf q}, 
\label{alphaA} 
\\ \alpha_B & = & \alpha_A e^{i({\bf R}_B - {\bf R}_A) \cdot {\bf q}} 
\equiv \alpha_A e^{i \eta_B},
\label{alphaB} 
\\ \alpha_C & = & \alpha_A e^{i({\bf R}_C - {\bf R}_A) \cdot {\bf q}}
\equiv \alpha_A e^{i \eta_C}
\label{alphaC}
\end{eqnarray}
where the last two equations define the phases $\eta_B$ and
$\eta_C$. This completes the specification of the qubit-bath coupling.


\subsection{The Redfield equation}
\label{sec:redfield}

We now investigate the qubit decoherence due to the bosonic bath by
determining the time relaxation of the system's reduced density
matrix. We use the Born and Markov approximations and the Redfield
equation.\cite{argyres64} In this formalism the reduced density matrix
of the system (qubit) is obtained by integrating out the bath degrees
of freedom and assuming that: (i) the coupling to the bath is weak, so
leading order perturbation theory is applicable (the Born
approximation), and (ii) the bath correlation time is much shorter
than the typical time scale of operation of the qubit, so that
system-bath interaction events are uncorrelated in time (the Markov
approximation).

The time evolution of the reduced density matrix is given by the
Redfield equation \cite{argyres64,pollard94}
\bea
\label{redfield}
{\dot{\rho}}(t) & =& - i\, [\tilde{H}_S(t), \rho(t)]\\ && {} +
\sum_{\alpha=1,2} \big\{ \left[ \Lambda_{\alpha}(t) \rho(t),
K_{\alpha} \right] + \left[ K_{\alpha}, \Lambda^{\dag}_{\alpha}(t)
\rho(t) \right] \big\}, \nonumber \eea
where the time-dependent auxiliary matrices $\Lambda_\alpha(t)$ which
encode the bath correlation properties are defined by
\begin{equation}
\Lambda_{\alpha}(t) = \sum_{\beta=1,2} \int_0^{\infty} dt^\prime
{B}_{\alpha \beta}(t^\prime)\, e^{-i t^\prime \tilde{H}_S(t)}
K_{\beta}\, e^{i t^\prime \tilde{H}_S(t)}.
\end{equation}
The thermal-average bath correlation functions,
\begin{equation}
{B}_{\alpha\beta}(t) = \langle \Phi_{\beta}(t)\, \Phi_{\alpha}(0)
\rangle,
\end{equation}
can be written in terms of spectral functions,
\begin{equation}
\label{eq:spec}
\nu_{\alpha\beta} (\omega) = \sum_{\bf q} g^{(\alpha)}_{\bf q}
g^{(\beta)}_{\bf -q} \delta (\omega - \omega_{\bf q}),
\end{equation}
and the boson occupation number $n_B(\omega) = (e^{\omega/T}
-1)^{-1}$:
\begin{eqnarray}
\label{bathcorr}
{B}_{\alpha\beta}(t) & = & \int_0^{\infty} d \omega\, \nu_{\alpha
             \beta} (\omega) \\ & & \times \left\{ e^{i \omega t}
             n_B(\omega) + e^{-i \omega t} \left[ 1 + n_B(\omega)
             \right]) \right\} \;.  \nonumber
\end{eqnarray}
Performing the sum over ${\bf q}$ in Eq.~(\ref{eq:spec}), we find
\begin{eqnarray}
\label{eq:specfunc11}
\nu_{11} & = & 2 \sum_{\bf q} \left| \lambda_{\bf q} P_{\bf q}
\right|^2 \delta (\omega - \omega_{\bf q} ) \nonumber \\ & & \times
\left[3 - 2( \cos\eta_B - \cos \eta_C) + \cos
(\eta_B-\eta_C)\right],\\
\label{eq:specfunc22}
\nu_{22} &=& 2 \sum_{\bf q} \left| \lambda_{\bf q} P_{\bf q} \right|^2
\delta (\omega - \omega_{\bf q} )\, [1-\cos (\eta_B - \eta_C)], \\
\label{eq:specfunc12}
\nu_{12} & = & 2 \sum_{\bf q} \left| \lambda_{\bf q} P_{\bf q}
\right|^2 \delta (\omega - \omega_{\bf q} ) \nonumber \\ & & \times
\left[ e^{-i\eta_B} - e^{-i\eta_C} + i \sin (\eta_B-\eta_C) \right],
\end{eqnarray}
with $\nu_{21} = \nu_{12}^\ast$. When the bath is sufficiently large,
the sums over the vector ${\bf q}$ in
Eqs.~(\ref{eq:specfunc11})-(\ref{eq:specfunc12}) can be converted into
three-dimensional integrals.

A few simplifying but realistic assumptions can be made at this
point. Let us first assume that the coupling constant $\lambda_{\bf
q}$ and the dispersion relation $\omega_{\bf q}$ are both
isotropic. Second, let us assume that the electronic density in the
dots has a Gaussian profile, $\rho({\bf r}) = \delta(z)\,
e^{-r^2/(2\,a^2)}/(2 \pi a^2)$, resulting in
\begin{equation}
\label{eq:formfactor}
P_{\bf q} = \int d^3r \, \rho(r)\, e^{-i {\bf q} \cdot {\bf r}} =
e^{-(aq\sin\theta)^2/2},
\end{equation}
where $(q,\theta,\varphi)$ are the spherical coordinates of the boson
wave vector. Then, the three-fold symmetry in the plane causes
$\nu_{12} (\omega)$ to vanish and
\begin{equation}
\nu_{11} (\omega) = 3\, \nu(\omega) \;, \quad
\nu_{22} (\omega) = \nu(\omega),
\end{equation}
with
\begin{eqnarray}
\nu(\omega) & = & \frac{\Omega\, q^2}{2\pi^2} |\lambda_q|^2 \left|
\frac{d\omega_q}{dq} \right|^{-1} \int_0^{\pi/2} d\theta\,
\sin\theta\, e^{-(qa\sin\theta)^2} \nonumber \\ & & \times \left[ 1
- J_0 \left( qD \sin\theta \right) \right],
\end{eqnarray}
where $\omega = \omega_q$, $\Omega$ is the crystal unit cell volume,
and $D$ is the distance between dots.

For III-V semiconductor materials at low temperatures, the most
relevant bosonic modes are piezoelectric acoustic phonons,\cite{bruus}
for which we have $\lambda_q = \pi s \sqrt{g_{\rm ph}/q\Omega}$ and
$\omega_q = sq$. Here, $g_{\rm ph}$ is the dimensionless
electron-phonon coupling constant and $s$ is the phonon velocity (for
GaAs, $g_{\rm ph} \approx 0.05$ and $s \approx 5 \times 10^{3}$
m/s).\cite{sergueietaldoubledot05}

\subsection{Decoherence rates}
\label{sec:decoherence}

We now solve the equation-of-motion for the reduced density matrix
explicitly for a case in which the decoherence rate can be obtained
directly. Consider a constant pulse applied to the qubit at $t \!=\!
0$ such that $h_y \!=\! h_z \!=\! 0$ and $h_x \!=\! \Delta > 0$. For
$t \!>\! 0$, the $\Lambda_\alpha$ matrices are constant and given by
\bea 
\Lambda_1 & = & \gamma_0\, \sigma_1/2 \;, \\ 
\Lambda_2 & = & - (1/2\sqrt{3})\, \left( \gamma_c\, \sigma_2 + \gamma_s\,
\sigma_3 \right).
\eea 
The (complex) relaxation rates are given by
\bea
\label{eq:gamma0}
\gamma_0 & \equiv & \int_0^\infty \!\!dt\, B_{22}(t),\\
\label{eq:gammac}
\gamma_c & \equiv & \int_0^\infty \!\!dt\, B_{22}(t)\, \cos(2\Delta t),\\
\label{eq:gammas}
\gamma_s & \equiv & \int_0^\infty \!\!dt\, B_{22}(t)\, \sin(2\Delta t).
\eea
The relaxation part of Eq.~(\ref{redfield}) then reads
\begin{widetext}
\begin{eqnarray}
\label{eq:redfield}
\sum_{\alpha=1,2} \big\{ [\Lambda_\alpha\, \rho , K_\alpha] + {\rm
h.c.}\big\} & = & \frac{\gamma_0^{\,\prime}}{6} \left(
\begin{array}{cc} 
\rho_{22} - \rho_{11} & \;\; \rho_{12}^\ast - \rho_{12} \\ \rho_{12} -
 \rho_{12}^\ast & \;\; \rho_{11} - \rho_{22}
\end{array}
 \right) + \frac{\gamma_c^{\,\prime}}{6} \left(
 \begin{array}{cc} 
\rho_{22} - \rho_{11} & \;\; -\rho_{12}^\ast - \rho_{12} \\ -\rho_{12}
- \rho_{12}^\ast & \;\; \rho_{11} - \rho_{22}
\end{array}
 \right) \nonumber \\ & & {} + i \frac{\gamma_s^{\,\prime}}{6} \left(
 \begin{array}{cc} \rho_{12} - \rho_{12}^\ast & 0  \\
 0 & \rho_{12}^\ast - \rho_{12} \end{array} \right) +
 \frac{\gamma_s^{\prime\prime}}{6} \left( \begin{array}{cc} 0 & 1 \\ 1
 & 0
 \end{array} \right),
\end{eqnarray}
\end{widetext}
where the single and double primes denote real and imaginary parts,
respectively. They can be easily evaluated, yielding
\bea \gamma_0^\prime & = & 0, \\ \gamma_c^\prime =
\gamma_s^{\prime\prime} & = & \frac{\pi}{2}\, \nu(2\Delta)\, {\rm
coth} \left( \frac{\Delta}{T} \right), \\ \gamma_s^\prime & = & -
\dashint_{0}^{\infty} \frac{dy}{y^2-1}\, \nu(2\Delta y)\, {\rm coth}
\left( \frac{\Delta y}{T} \right).  \eea
The Liouville term in Eq.~(\ref{redfield}) is obtained
straightforwardly:
\begin{equation}
\label{eq:liouville}
-i\, [ \tilde{H}_S,\rho] = -i\Delta\, [\sigma_1,\rho] = -i\Delta\,
 \left( \begin{array}{cc} \rho_{12}^\ast - \rho_{12} & \; \rho_{22} -
 \rho_{11} \\ \rho_{11} - \rho_{22} & \; \rho_{12} -\rho_{12}^\ast
\end{array} \right).
\end{equation}
Introducing Eqs.~(\ref{eq:redfield}) and (\ref{eq:liouville}) into
(\ref{redfield}), we obtain
\begin{eqnarray}
\label{firsteq}
\dot{\rho}_{11} & = & - 2 \left( \Delta +
\frac{\gamma_s^{\,\prime}}{6} \right) \, \rho_{12}^{\prime\prime} +
\frac{\gamma_c^{\,\prime}}{6}\, (1 - 2\rho_{11}), \\
\label{thirdeq}
 \dot{\rho}^{\,\prime}_{12} & = & -
\frac{\gamma_c^{\,\prime}}{3} \, \rho^{\,\prime}_{12} +
\frac{\gamma_c^\prime}{6},\\
\label{secondeq}
\dot{\rho}^{\prime\prime}_{12} & = & - \Delta \, (1 - 2\rho_{11}),
\end{eqnarray}
where we have split the off-diagonal term $\rho_{12}$ into real and
imaginary parts, $\rho_{12}^{\,\prime} + i \rho_{12}^{\prime \prime}$.

In order to identify energy and phase relaxation rates, we rewrite the
elements of the reduced matrix in the eigenbasis of the system
Hamiltonian,
\begin{equation}
|E=\pm \Delta \rangle = \frac{1}{\sqrt{2}} \left( |+\rangle \pm
 |-\rangle \right),
\end{equation}
resulting in
\begin{eqnarray}
\dot{\tilde{\rho}}_{11} & = & - \frac{\gamma_c^\prime}{3}
\tilde{\rho}_{11} + \frac{\gamma_c^\prime}{3},\\
\dot{\tilde{\rho}}^{\,\prime}_{12} & = & - \left( 2\Delta +
\frac{\gamma_s^\prime}{3} \right) \tilde{\rho}^{\,\prime\prime}_{12} -
\frac{\gamma_c^\prime}{3} \tilde{\rho}^{\,\prime}_{12},\\
\dot{\tilde{\rho}}^{\,\prime\prime}_{12} & = &
2b\,\tilde{\rho}_{12}^{\,\prime}.
\end{eqnarray}
The solution of the diagonal term is straightforward,
\begin{equation}
\tilde{\rho}_{11}^{\,\prime}(t) = 1 + \left[
\tilde{\rho}_{11}^{\,\prime} (0) - 1 \right]\, e^{-\gamma_c^\prime
t/3},
\end{equation}
which allows us to read directly the energy relaxation time,
\begin{equation}
\label{eq:T1}
T_1 = \frac{3}{\gamma_c^{\,\prime}}.
\end{equation}
For the off-diagonal term, one finds that the real part is given by
\begin{equation}
\tilde{\rho}_{12}^{\,\prime}(t) = \tilde{\rho}_{12}^{\,\prime}(0)\,
e^{-t/T_2}\, \cos(\omega_c t),
\end{equation}
where the phase relaxation time is equal to
\begin{equation}
\label{eq:T2}
T_2 = \frac{6}{\gamma_c^\prime}
\end{equation}
and the frequency of quantum oscillations is given by
\begin{equation}
\label{eq:omega_c}
\omega_c = \sqrt{2\Delta \left( 2\Delta + \frac{\gamma_s^\prime}{3}
\right) - \frac{\gamma_c^{\prime\,2}}{36}}.
\end{equation}
Note that $T_2 = 2T_1$, as well-known for the super-ohmic 
spin-boson model in the weak-coupling regime.\cite{slichter,weiss} 

\begin{figure}[t]
\includegraphics[width=8cm]{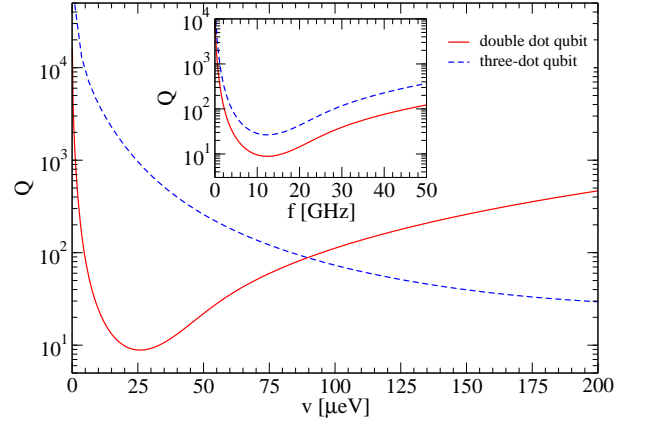}
\caption{(Color online) Comparison between the $Q$ factors of a
three-dot and a double-dot charge qubit coupled to piezoelectric
acoustic phonons. The parameters used are: $a = 60$ nm, $D=180$ nm,
$s=5\times 10^3$ m/s, $T=15$ mK, and $g_{\rm ph} = 0.05$, which
correspond to realistic lateral quantum dot systems in GaAs. Here the
variable $v$ denotes the interdot tunnel amplitude. Note that for
double dot qubits, $\Delta = v$, while for three-dot qubits we assumed
$\Delta = 0.1\, v$. In the inset we show the same $Q$ factors when the
oscillation frequency (rather than $v$) is fixed. In this case the
curves only differ by a factor of 3.}
\label{fig:Qfac3x2}
\end{figure}

Except for a factor of three in the relaxation rates, Eqs.
(\ref{eq:T1})-(\ref{eq:omega_c}) are identical to those found in
Ref.\,\onlinecite{sergueietaldoubledot05} for a double-dot charge
qubit. However, one has to recall that while in the double-dot qubit
$\Delta$ is the interdot hopping matrix element $v$, for the three-dot
qubit it takes a much smaller value, of the order of
$\delta_{1,2,3}$. The decoherence times will be longer for the
three-dot qubit, but so will be the quantum oscillation period and the
single-qubit gate pulses. Therefore, it is meaningful to compare the
quality factor of the the three-dot qubit to that obtained for the
double-dot qubit for a {\it fixed magnitude} of $v$, which is a common
experimental parameter to both setups. The comparison for the case of
piezoelectric acoustic phonons and realistic GaAs quantum dot
geometries (data for the double-dot qubit was obtained from
Ref.\,\onlinecite{sergueietaldoubledot05}) is shown in
Fig.~\ref{fig:Qfac3x2}. The $Q$ factor is defined as
\begin{equation}
Q = \frac{\omega_c\, T_2}{2\pi}.
\end{equation}
We assume $\omega_c \approx 2\Delta$ since $\Delta \gg
\gamma^\prime_c,\gamma^\prime_s$ in the weak-coupling regime. 



\textit{The improvement in the $Q$ factor is substantial for small
tunnel amplitudes.} A similar result was previously found by Storcz
{\it et al.} in Ref.\,\onlinecite{storcz05} when considering the
phonon-induced decoherence in a system of two double-dot charge qubits
with a small tunnel splitting (``slow tunneling''). There, the
dominant quadrupolar contribution to the {\it two-qubit} decoherence
yields a $\omega_c^5$ dependence for the $Q$ factor. In our case, the
extra protection in the three-dot qubit compared to the slow tunneling
double-dot system arises mainly because the oscillation frequency
$\omega_c$ (i.e., the amplitude of the transverse pseudomagnetic
field) is smaller in the three-dot qubit by the ratio $\Delta/v$ [see
Eq.~(\ref{eq:Hamilton})]. This ratio must be kept small in order to
avoid leakage from the computational basis. In Fig.~\ref{fig:Qfac3x2}
it was set to 0.1. However, for a fixed oscillation frequency (see
inset in Fig.~\ref{fig:Qfac3x2}), the $Q$ factors for these two qubits
differ by only an overall factor of three.

To summarize up to this point, our study indicates that using a
computational basis with a homogeneous charge distribution improves
the quality of the qubit but does not rid it from decoherence
completely. The reason lies in the fact that bosonic modes propagating
in the $xy$ plane can pick up distinct phase shifts when interacting
with different dots [see Eqs.~(\ref{alphaA}) to
(\ref{alphaC})]. However, there is no complete destructive
interference along any direction of propagation in the plane, as can
be seen from Eqs.~(\ref{eq:g1}) and (\ref{eq:g2}). In fact, one can
show that the same is true for any ringlike array of dots that share a
single excess electron.

\section{Charge qubits in multipole configurations}
\label{sec:multipledots}

\begin{figure}[t]
\includegraphics[width=7cm]{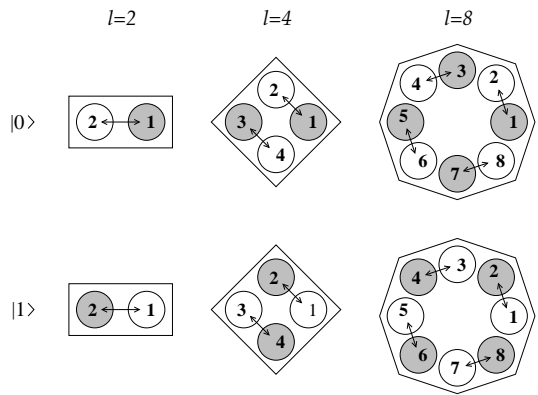}
\caption{The three lowest multipole charge qubit configurations
(dipole, quadrupole, and octopole). The two computational basis
states, $|0\rangle$ and $|1\rangle$, are indicated for for each
configuration. Empty (filled) circles correspond to empty (occupied)
quantum dots. The arrows indicate the pairs of quantum dots where
excess charge can hop.}
\label{fig:multp}
\end{figure}

As recently proposed by Oi {\it et al.},\cite{oi05} there is another
way in which the geometry of the quantum dot qubit array and its
charge distribution can be chosen to minimize the coupling to
environmental degrees of freedom. Here we demonstrate how their idea
can be extended to multiple-dot charge qubits coupled to gapless
bosonic modes. It turns out that by reducing the computational space
to particular multipole charge configurations one can substantially
reduce the coupling to bath modes at low frequencies. We consider
qubits and basis states as shown in Fig.~\ref{fig:multp}. The qubit
consists of a planar array of dots with alternating excess
charge. Note that the operation of such a qubit is straightforward:
The excess charge is only allowed to hop between every other pair of
neighboring dots, namely, between dots numbered $2n-1$ and $2n$, with
$n=1,\ldots, 2^{p-1}$, where $p$ is the multipole order, $l=2^p$ (see
Fig.~\ref{fig:multp}). Tunnel barriers between alternating pairs of
dots must be maintained small and fixed (to avoid leakage), while the
remaining barriers have to be modulated in time to implement an $X$
gate. The $Z$ gate is implemented by inducing a small bias between
even- and odd-numbered dots. Two-qubit operations can be implemented
in analogy to the procedure discussed in Ref.\,\onlinecite{nielsen00}
in the context of Ising-interaction based two-qubit operations. 

The basis states for each multipole configuration have complementary
charge distributions that tend to cancel out the coupling to phonon
modes propagating along certain directions in the $xy$ plane. The
number of such directions increases with the multipole order,
resulting in an attenuation of the overall coupling to phonons at low
frequencies (large wavelengths). The crossover frequency where this
attenuation occurs is $\omega_{\rm cross}^{(l)} \sim s/d_l$, where
$d_l$ is the radius of the dot array. At high frequencies, however,
when the phonon wavelength is much smaller than the radius $d_l$,
decoherence becomes stronger because phonons can resolve the internal
structure of the qubit and disturb charge motion between individual
pairs of dots.

In order to demonstrate these effects, let us derive an expression for
the spectral function of the qubit-bath system. For simplicity, we
assume that all dots in the qubit are identical. In this case, the
bath modes couple to charge variations in the dots according to the
Hamiltonian
\begin{equation}
H_{\rm SB} = \sum_{\bf q} \sum_{k=1}^l \alpha_k\, N_k \, (b_{\bf
q}^\dagger + b_{\bf -q}),
\end{equation}
where $l=2^p$ and $N_k$ is the excess charge in the $k$th dot. For the
case of acoustic phonons, the coefficients $\alpha_k$ were defined in
Eq.~(\ref{eq:alpha}). Projecting this Hamiltonian onto the
computational basis (as shown in Fig.~\ref{fig:multp}), we find that,
up to a constant term,
\begin{equation}
H_{\rm SB} = K\, \Phi_l,
\end{equation}
where $K = -\sigma_z/2$ acts on the qubit space and
\begin{equation}
\Phi_l = \sum_{\bf q} g^{(l)}_{\bf q} (b_{\bf q}^\dagger + b_{\bf
-q}),
\end{equation}
acts on the phonon bath, with
\begin{equation}
g^{(l)}_{\bf q} = \sum_{k=1}^l (-1)^k \alpha_k = \lambda_{\bf q}\,
P_{\bf q}\, \sum_{k=1}^l (-1)^k e^{i{\bf R}_k \cdot {\bf q}}.
\end{equation}
It is convenient to choose the position vectors of the dots as ${\bf
R}_k = d_l( \hat{x}\, \cos\varphi_k + \hat{y}\,\sin\varphi_k)$, where
$\varphi_k = (2\pi/l)(k-1)$ and $d_l$ is the array radius: $d_l =
D/2\sin(\pi/l)$, where $D$ is the distance between neighboring
dots. This yields
%
\begin{eqnarray}
\lefteqn{ \left| g^{(l)}_{\bf q} \right|^2 = |\lambda_{\bf q}\, P_{\bf
q}|^2\, \sum_{k,j=1}^l (-1)^{k+j} } \\ & & \times \exp\! \Big[ 2id_l
q\sin\theta \sin \Big( \varphi - \frac{\varphi_k+\varphi_j}{2} \Big)
\sin\Big( \frac{\varphi_k-\varphi_j}{2} \Big) \Big], \nonumber
\end{eqnarray}
where $(q,\theta,\varphi)$ are the spherical coordinates of the wave
vector ${\bf q}$. It is not difficult to see that $g_{\bf q} = 0$ for
$\theta=\pi/2$ and $\varphi = (2m-1)\pi/l$, with $m=1,\ldots,l$.

The spectral function can now be obtained in analogy to the
calculation shown in Sec.~\ref{sec:redfield}. For a thermal bath of
acoustic piezoelectric phonons, we find
%
\begin{eqnarray}
\label{eq:nul}
\nu_l(\omega) & = & \sum_{\bf q} \left| g^{(l)}_{\bf q} \right|^2\,
\delta(\omega - \omega_{\bf q}) \nonumber \\ & = & \frac{g_{\rm ph}\,
\omega\, l}{2} \int_0^{\pi/2} \!d\theta\, \sin\theta\, \exp\! \left( -
\frac{a^2\omega^2\sin^2\theta}{s^2} \right) \nonumber \\ & & \times
\Bigg\{ 1 + (-1)^{l/2} J_0 \left( \frac{2d_l\omega}{s} \sin\theta
\right) \\ & & + 2 \sum_{m=1}^{l/2-1} (-1)^m J_0 \left[
\frac{2d_l\omega}{s} \sin\theta \sin \left( \frac{m\pi}{l}
\right)\right] \Bigg\}. \nonumber
\end{eqnarray}
%
Implicit in Eq.~(\ref{eq:nul}) are the assumptions of in-plane
isotropy of $\lambda_{\bf q}$, $P_{\bf q}$, and $\omega_{\bf q}$. Note
that for $l=2$ one recovers the spectral function for a double dot
qubit obtained in Ref.\,\onlinecite{sergueietaldoubledot05}. The
low-frequency behavior of the spectral density becomes apparent when
we expand the Bessel functions in a power series, resulting in
\begin{eqnarray}
\label{eq:nulg}
\nu_l(\omega) & = & \frac{g_{\rm ph}\, \omega\, l}{2} \int_0^{\pi/2}
d\theta\, \sin\theta\, \exp \left( - \frac{\omega^2a^2}{s^2}
\sin^2\theta \right) \nonumber \\ & & \times \sum_{k=1}^\infty
\frac{(-1)^k}{(k!)^2}\, \left( \frac{d_l\omega}{s} \sin\theta
\right)^{2k} a_k^{(l)},
\end{eqnarray}
where
\begin{equation}
a_k^{(l)} = (-1)^{l/2} + 2\sum_{m=1}^{l/2-1} (-1)^m \sin^{2k} \left(
\frac{m\pi}{l} \right).
\end{equation}
It is possible to show that $a_k^{(l)} = 0$ for $k < l/2$ when $l$ is
an integer power of 2. Therefore, $\nu_l(\omega\rightarrow 0) \sim
\omega^{l+1}$. For large $l$, this amounts to the appearance of a
pseudo gap in the spectral function at low frequencies. The asymptotic
behavior of the spectral function at high frequencies is also
straightforward to obtain: One finds $\nu_l(\omega \rightarrow \infty)
\sim l/\omega$. Thus, the tail of the spectral function raises with
increasing multipole order.

\begin{figure}[b]
\includegraphics[width=7cm]{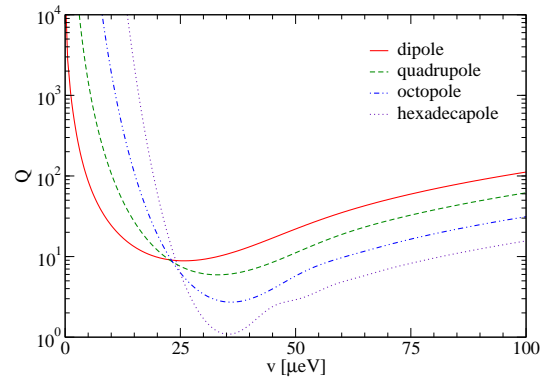}
\caption{(Color online) $Q$ factors for multipole charge qubits
($l=2,4,8,16$) coupled to piezoelectric acoustic phonons: $Q_l =
\omega_c/\pi\gamma^{(l)}$, where $\omega_c \approx 2v$ [for
$\gamma^{(l)}$, see Eq.~(\ref{eq:gammal})]. Physical and geometrical
parameters are the same as those used in Fig.~\ref{fig:Qfac3x2}. In
particular, note that the inter-dot distance is fixed, $D = 120$ nm,
for all configurations.}
\label{fig:Qfac_mult}
\end{figure}

The structure of the computational basis is simple enough to allow for
the qubit to couple to just one bath mode (in contrast to the
three-dot qubit, where two modes couple to the qubit). Thus, the
standard expressions for the relaxation times in the spin-boson model
can be used.\cite{weiss} The result is
\begin{equation}
\label{eq:gammal}
\gamma^{(l)} = \frac{\pi}{2} \nu_l(2v)\, \mbox{coth} \left(
\frac{v}{T} \right),
\end{equation}
where $v$ is the inter-dot tunnel amplitude. Note that
Eq.~(\ref{eq:gammal}) reduces to the result found in
Ref.\,\onlinecite{sergueietaldoubledot05} when $l=2$, as
expected. Provided that $v$ is sufficiently smaller than the
temperature, $\gamma^{(l)} \sim v^l$. Thus, by increasing the order of
the multipole and maintaining a low frequency of operation, one can
decrease the qubit relaxation rate by orders of magnitude without
affecting the frequency of quantum oscillations. In
Fig.~\ref{fig:Qfac_mult} we show the $Q$ factor of several multipole
charge qubits as a function of the inter-dot coupling $v$. Note that
at low frequencies high quantum oscillations are much less damped for
high multipole configurations. This translates into single-qubit gates
of much higher fidelity. Clearly, this gain in the $Q$ factor has to
be contrasted with the high complexity of operating a logical qubit
comprised by a large number of quantum dots, as well with the slowness
in operation. As the gating becomes slower, other sources of
decoherence may become more relevant.

It is also important to remark that, in practice, the pseudogap width,
$\omega_{\rm cross}^{(l)}$, will shrink with increasing multipole
order. This is because the dot array radius scales as $d_l \approx l\,
D/2\pi$ for $l\gg 1$. Therefore, for a fixed value of $D$, one has
$\omega_{\rm cross}^{(l)} \sim 2\pi\, s/(l\, D)$ for $l\gg
1$. Finally, we note that the results discussed above are valid for
any gapless bosonic bath. Different dependences on $q$ for the
coupling constant $\lambda_q$ and dispersion relation $\omega_q$ will
only change the power of the frequency-dependent prefactor in
Eq.~(\ref{eq:nulg}).

\section{Conclusions}
\label{sec:conclusions}


%

In this paper we have shown that, whereas there are no simple ways to
completely protect charge qubits based on quantum dots from
decoherence by gapless bosonic modes, a homogeneous charge
distribution throughout the qubit is the most advantageous setup and
provides the best possible protection against decoherence. This result
applies not only to the charge qubits in
semiconductor-heterostructures that we focused on here, but, in
principle, to charge qubits in general. Whereas certain aspects of the
discussion need to be changed for, say, self-assembled quantum dots,
single-donor charge qubits,\cite{hollenberg04} or Si-based quantum dot
structures,\cite{gorman05} this does not affect the universal
mechanism underlying our central result, namely that a specific
(homogeneous) charge distribution within the qubit enables the
cancellation of (certain) decoherence modes.

Contrary to spin-based quantum dot qubits, where decoherence-free
subspaces can be created by combining quantum dots into logical units,
charge-based qubits are much more difficult to isolate from the
environment. In order to have decoherence-free subspaces for charge
qubits one would need to restrict the operation to a subspace where
charge is homogeneously distributed in space, no matter which basis
states are chosen. However, this contradicts the very nature of a
charge qubit (where readout depends on charge imbalance) and thus
cannot be achieved. In our example of the three-dot qubit these facts
become evident in the existence of two phonon modes that {\it cannot
cancel due to geometric constraints} inherent to the qubit.

Decoherence can be mitigated in a number of other ways. For instance,
for the three-dot qubit case we have studied, a substantial
improvement with respect to the double-dot qubit can be achieved due
to the lower frequency of operation and to an enhancement of the
relaxation time by a factor of three.

Another effective way to reduce the coupling to gapless bosonic modes
is to choose a computational basis with a multipole charge
configuration. As we have shown, the multipole geometry attenuates the
coupling to long wavelength acoustic phonons by a factor proportional
to a power law of the operation frequency. This power law grows
rapidly with the multipole order. Thus, multipole configurations of
charge can lead to quality factors enhanced by orders of magnitude in
comparison to those obtained for double-dot qubits. However, the
effect is reversed at high frequencies of operation. The crossover
frequency separating the two regimes is given by the inverse traversal
time for a phonon to propagate across the qubit. For typical GaAs
setups, this time is of the order of 30 ps (for dots 120 nm apart),
indicating a crossover frequency in the range of 30 GHz. Since tunnel
amplitudes usually vary from tens to a few hundreds of $\mu$eV,
yielding quantum oscillations of about 2-20 GHz, there is a real
advantage in moving toward multiple-dot qubits for current
setups. However, since phonons are not the only leading mechanism for
decoherence in charge qubits,\cite{sergueietaldoubledot05} as
operation frequencies go down other sources of decoherence, not
necessarily modeled by bosonic environments, may become dominant. In
that case multiple-dot qubits might become less
appealing. Investigation of other decoherence mechanisms in
multiple-dot qubits, such as electromagnetic environmental
fluctuations and telegraph noise due to charges trapped in the
substrate, are the subject of ongoing investigations.

Finally, we mention that a recent work has shown that pulse
optimization is also a very effective way of minimizing the coupling
to bosonic environments in quantum dot charge
qubits.\cite{hohenester07}

\section{Acknowledgments}

We would like to thank J. Siewert for useful discussions. M.H. thanks
the Alexander von Humboldt Foundation for partial support and also
acknowledges funding through the Emmy-Noether program of the German
Research Foundation (DFG). This work was also supported in part by the
NSA and ARDA under ARO Contract No. DAAD19-02-1-0079 and by NSF Grants
No. CCF-0523509 and No. CCF-0523603. D.C.B.V. and E.R.M. acknowledge
partial support from the Interdisciplinary Information Science and
Technology Laboratory (I$^2$Lab) at UCF.

\appendix

\section{THE TWO-QUBIT REDUCED HAMILTONIAN}
\label{ReducedHamAppendix}

The Hamiltonian of two three-dot qubits coupled by their bases [see
Fig.~\ref{fig_2qubitoper}(b)] with inter-qubit couplings $t^\prime$
and $t^{\prime\prime}$ has the following matrix form in the basis of
Eqs.~(\ref{eq:1})-(\ref{2qubitbasis10}) (the lower off-diagonal block
is omitted):
\begin{widetext}
\begin{equation}
\label{eq:H2qubit}
H_{\rm I-II}^{\rm base} = \left(
\begin{array}{cccccccccc}
0 & 0 & 0 & 0 & -\tpm & -\tppn & \tpn - \tppm & \tppm & \tpn & \tpm -
\tppn \\ 0 & 0 & 0 & 0 & -\tpn & \tppm & t^\prime/3 -
t^{\prime\prime}/3 & \tppm & \tpn & t^\prime/3 - t^{\prime\prime}/3 \\
0 & 0 & 0 & 0 & -\tpm & -\tppn & t^\prime/3 - t^{\prime\prime}/3 &
\tppn & \tpm & t^\prime/3 - t^{\prime\prime}/3 \\ 0 & 0 & 0 & 0 &
-\tpn & -\tppm & \tpm -\tppn & \tppn & \tpm & \tpn - \tppm \\ & & & &
U_i &v & -v & 0 & 0 & 0 \\ & & & & v & U_i& v & 0 & 0 & 0 \\
& & & & -v & v & U_i & 0 & 0 & 0 \\ & & & & 0 & 0 & 0 & U_i
& v & -v \\ & & & & 0 & 0 & 0 & v & U_i & v \\ & & & & 0 & 0 & 0
& -v & v & U_i
\end{array} \right).
\end{equation}
\end{widetext}
[Note that the Hamiltonian for the tip-tip configuration is recovered
by setting $t^{\prime\prime} = 0$ in Eq.~(\ref{eq:H2qubit}).] This
Hamiltonian can be projected onto the two-qubit computational subspace
by means of a Schrieffer-Wolff transformation. From
Eq.~(\ref{eq:H2qubit}), we see that the Hamiltonian has the form
$H_{\rm I-II}^{\rm base} = H_0 + H_1$, where
\begin{equation}
H_0 = \left( \begin{array}{cc} 0 & 0 \\ 0 & M \end{array} \right),
\quad H_1 = \left( \begin{array}{cc} 0 & T \\ T^\dagger & 0
\end{array} \right),
\end{equation}
and $M$ and $T$ are $6 \times 6$ and $4 \times 6$ matrices,
respectively. Performing the Schrieffer-Wolff transformation and
expanding to second order in $H_1$,\cite{schriefferwolfftrafo} we get
$\tilde{H}_{\rm I-II}^{\rm base} \approx H_0 + (1/2) [S,H_1]$, where
\begin{equation}
S = \left( \begin{array}{cc} 0 & -T\, M^{-1} \\ -M^{-1}\, T^\dagger &
0
\end{array} \right).
\end{equation}
Thus the Hamiltonian has the block diagonal structure
\begin{equation}
\tilde{H}_{\rm I-II}^{\rm base} \approx \left( \begin{array}{cc}
H^{\rm red}_{\rm I-II} & 0 \\ 0 & M
\end{array} \right),
\end{equation}
where $H^{\rm red}_{\rm I-II} = -T\, M^{-1}\, T^\dagger$. The matrix
$M$ can be broken into two identical $3 \times 3$ diagonal blocks,
\begin{equation}
M = \left( \begin{array}{cc} B & 0 \\ 0 & B
\end{array} \right),
\end{equation}
and $T$ can be broken into two distinct $4 \times 3$ blocks,
\begin{equation}
T = \left( \begin{array}{cc} T_{\rm I} & T_{\rm II}
\end{array} \right).
\end{equation}
As a result, $H^{\rm red}_{\rm I-II} = -T_{\rm I}\, B^{-1}\, T_{\rm
I}^\dagger - T_{\rm II}\, B^{-1}\, T_{\rm II}^\dagger$. After some
algebra, one finds that
\begin{equation}
B^{-1} = \frac{1}{3}\, \left( \begin{array}{ccc}
2 u_1 + u_2 & u_1 - u_2 & -u_1 + u_2 \\
u_1 - u_2 &  2 u_1 + u_2 & u_1 - u_2 \\
-u_1 + u_2 & u_1 - u_2 & 2 u_1 + u_2
\end{array} \right),
\end{equation}
where $u_1 = (U_i + v)^{-1}$ and $u_2 = (U_i - 2 v) ^{-1}$. The
structure of $B^{-1}$ can be substantially simplified by assuming $v
\!\ll\! U_i$ and neglecting $v$. This yields $H^{\rm red}_{\rm I-II} =
(-1/U_i) \left( T_{\rm I}^{}\, T_{\rm I}^\dagger + T_{\rm II}\, T_{\rm
II}^\dagger \right)$. Carrying out the matrix multiplications and
setting $t^{\prime\prime} = 0$, we obtain Eq.~(\ref{eq:Hreduced1}).




\begin{thebibliography}{99}
\bibitem{note_holestates}
By working with 1 hole per three-dot qubit (ie. 5 electrons in 3 levels) instead of 1 electron, the degeneracy between eigenstates $|+\rangle$ and
$|-\rangle$ occurs at $B \!=\! 0$, which may have some advantages.
\bibitem{technicalnote} A more technical argument for neglecting the
states $|T\rangle_{\rm I}$ and $|T\rangle_{\rm II}$ can be constructed
as follows. First, keep all such states while performing a first
Schrieffer-Wolff transformation to find the effect of virtual
excitation to the doubly-occupied states. This yields an effective
Hamiltonian in the space of singly-occupied states with magnitude of
order $t^{\prime\, 2}/U_i$. It includes off-diagonal terms between the
computational basis and the $|T\rangle$ states. Now, perform a second
Schrieffer-Wolff transformation relying on the fact that $t^{\prime\,
2}/U_i \ll v$ to integrate out the $|T\rangle$ states. The
contribution of this second transformation to the effective
Hamiltonian in the computational basis is of order $(t^{\prime\,
2}/U_i)^2/v$. As this is much smaller than $t^{\prime\, 2}/U_i$, the
effect of the $|T\rangle$ states may be safely neglected.
\end{thebibliography}
\end{document}